\documentclass[twocolumn,showpacs,preprintnumbers,amssymb]{revtex4}
\usepackage{pstricks}%
\usepackage{graphicx}
\usepackage{dcolumn}
\usepackage{bm}
\usepackage{dsfont}
\usepackage{amsmath}
\usepackage{pifont}
\usepackage{psfrag}
\usepackage[hypertex]{hyperref}
\usepackage[latin2]{inputenc}

\newcommand{\BE}{\begin{equation}}
\newcommand{\EE}{\end{equation}}

\newcommand{\skipc}[2]{}

\newcommand{\eq}[1]{Eq.~(\ref{#1})}

\newcommand{\I}{\ensuremath{{\mkern1mu\mathrm{i}\mkern1mu}}}
\newcommand{\E}{\ensuremath{{\mkern1mu\mathrm{e}\mkern1mu}}}

\begin{document}

\title{Higher order wave-particle duality}

\author{Jie-Hui Huang$^{1}$, Sabine W\"olk$^{2}$, Shi-Yao Zhu$^{1}$, and M. Suhail Zubairy$^{1,2}$}

\affiliation{$^1$Beijing Computational Science Research Center, Beijing, 100084,
People's Republic of China\\
$^2$Institute of Quantum Science and Engineering (IQSE) and
Department of Physics and Astronomy, Texas A$\&$M University,
College Station, TX 77843-4242, USA}

\date{\today}

\begin{abstract}
The complementarity of single-photon's particle-like and wave-like
behaviors can be described by the inequality $D^2+V^2 \leq 1$, with
$D$ being the path distinguishability and $V$ being the fringe visibility. In this
paper, we generalize this duality relation to multi-photon case,
where two new concepts, higher order distinguishability and higher
order fringe visibility, are introduced to quantify the higher order
particle-like and wave-like behaviors of multi-photons.
\end{abstract}


\maketitle

\section{Introduction\label{intro}}
The complementarity principle, developed and introduced by Bohr in
1927 \cite{bohr}, is fundamentally important in the quantum theory,
which predicts that a quantum system may exhibit different
properties based on different measurement schemes. As the most
typical example of complementarity, wave-particle duality has
attracted much attention since the early days of the quantum theory
\cite{complementarity}. By defining the particle-like knowledge of
an object governed by quantum mechanics as the distinguishability
($D$) of its passage in a two-path interferometer, and the wave-like
knowledge as the visibility ($V$) of the interference pattern behind
the interferometer, a tradeoff relation between an object's
particle-like and wave-like behaviors can be established
\cite{Wootters,Glauber,Yasin,Rempe,Englert},
\begin{align} \label{eq1}
D^2+V^2 \le 1,
\end{align}
where the equal sign holds for pure state of single-particles. This
duality relation, already confirmed in many experiments
\cite{experiments}, is valid even when the choice of measuring
apparatus is delayed after the single-particle's entrance into the
interferometer \cite{wheeler}. Such a delayed-choice gedanken
experiment has been realized in experiments at single-photon level
by Roch group \cite{Roch}. The scheme of quantum eraser
\cite{eraser}, with the experimental realization reported in
Ref.\cite{shiyh}, provides another clear way to demonstrate the
exclusive relation between an object's particle-like and wave-like
behaviors. Recently, a new optical device named quantum beam
splitter (QBS) is theoretically proposed in Ref.  \cite{qbs1,qbs2},
and the wave-particle morphing behaviors of single-photons in this
quantum device is becoming a hot topic \cite{qbs3}.

In this paper, we generalize the discussion on the duality of
single-photons and investigate the higher order duality relations of
multi-photons, no matter what kind of state, pure or mixed, is
prepared for the multi-photons.

The organization of the paper is as follows: In section \uppercase\expandafter{\romannumeral2} we introduce
two new concepts, higher order
distinguishability and visibility. In  section
\uppercase\expandafter{\romannumeral3}, we derive an inequality for higher order duality. In  section \uppercase\expandafter{\romannumeral4}, we present a physical
interpretation on the higher order distinguishability and visibility
and in section \uppercase\expandafter{\romannumeral5} we present a
measurement scheme for the higher order visibility.

\section{Higher order distinguishability and visibility}
Before the concept of higher order duality is introduced, we first
recall the definitions of the particle-like information and the
wave-like information for single-photons. By feeding the
interferometer with single-photons, the particle-like information is
usually quantified as the distinguishability ($D$) of
single-photons' passage along the two paths inside the
interferometer (see Fig. 1). Thus we can use an operator
\cite{shanxi},
\begin{align} \label{eq2}
\hat{D} \equiv \frac{a_1^\dagger a_1 - a_2^\dagger a_2}{\langle a_1^\dagger a_1\rangle +\langle a_2^\dagger a_2\rangle },
\end{align}
to describe the measurement of the particle-like information
mentioned above. Here $a_1^\dagger~(a_1)$ and $a_2^\dagger~(a_2)$
denote the creation (annihilation) operators of the modes in path
$1$ and $2$, and the denominator $\langle a_1^\dagger a_1\rangle
+\langle a_2^\dagger a_2\rangle$ is for normalization. The wave-like
information is defined as the visibility ($V$) of the interference
pattern after the single-photons pass through the A
interferometer, whose measurement is in accord with the following
operator,
\begin{align} \label{eq3}
\hat{V} \equiv\left. \frac{ a_1^\dagger a_2 \E^{\I\phi} +
a_2^\dagger a_1\E^{-\I\phi}  }{\langle a_1^\dagger a_1\rangle
+\langle a_2^\dagger a_2\rangle}\right.
\end{align}
By describing the distinguishability and  visibility in terms of
operators, the particle-like and wave-like information of
single-photons are then the modules of the two expectation values,
i.e., $D=|\langle \hat{D} \rangle|$ and $V=|\langle
\hat{V^\prime}\rangle|_{\text{max by } \phi}$, where the phase
parameter $\phi$, controlled by the phase shifter in the
interferometer, should be appropriately chosen to maximize the
expectation value of the operator (\ref{eq3}).

The terms $a_1^\dagger a_1$ ($a_2^\dagger a_2$) in Eq.
(\ref{eq2}) and $a_1^\dagger a_2$ in Eq. (\ref{eq3}) are just the
first order auto-correlation of the field in path $1$ ($2$) and the
first order coherence between the fields in the two paths
\cite{correlation}, respectively. Therefore the distinguishability and
visibility defined in Eqs. (\ref{eq2}) and (\ref{eq3}) can be
regarded as the difference and coherence between the first order
correlation function of the two fields in the two paths $1$ and $2$.

Based on this viewpoint, we now introduce the concepts of
$k$th-order distinguishability,
\begin{subequations} \label{eq5}
\begin{align} \label{eq5a}
\hat{D}_k\equiv \frac{ (a_1^\dagger)^k a_1^k- (a_2^\dagger)^k
a_2^k}{\langle (a_1^\dagger )^k a_1^k\rangle + \langle
(a_2^\dagger)^k a_2^k\rangle},
\end{align}
\text{and $k$th-order visibility,}
\begin{align} \label{eq5b}
\hat{V}_k\equiv  \frac{(a_1^\dagger)^k a_2^k \E^{\I k\phi} +
(a_2^\dagger)^k a_1^k\E^{-\I k\phi}  }{\langle (a_1^\dagger )^k
a_1^k\rangle + \langle (a_2^\dagger)^k a_2^k\rangle},
\end{align}
\end{subequations}
the denominator $\langle (a_1^\dagger )^k a_1^k\rangle + \langle
(a_2^\dagger)^k a_2^k\rangle$ in both equations is again for
normalization. Here we have used $k$th-order auto-correlation
$(a_1^\dagger)^k a_1^k$ ($(a_2^\dagger)^k a_2^k$) and $k$th-order
coherence $(a_1^\dagger)^k a_2^k$ to replace the first order
auto-correlation and the first order coherence used in Eqs.
(\ref{eq2}) and (\ref{eq3}), which can now be regarded as the
special case of the definitions in (\ref{eq5}) by setting $k=1$.
Similar to the above treatment on the first order distinguishability
and visibility, the $k$th-order particle-like information and
$k$th-order wave-like information are just the modules of the
corresponding expectation values, i.e., $D_k=|\langle \hat{D}_k
\rangle|$ and $V_k=|\langle \hat{V}_k\rangle|_{\text{max by }
\phi}$. Here the phase parameter $\phi$ should be appropriately
chosen to maximize the visibility $V_k$, whose measurement will be
introduced in section \uppercase\expandafter{\romannumeral5} in more
details.


\section{Inequality for higher order duality}
Now we define the $k$th-order duality as the sum of the squared
$k$th-order particle-like information and $k$th-order wave-like
information, which is,
\begin{align} \label{eq6}
D_k^2+ V_k^2 =& \left(\frac{\langle (a_1^\dagger)^k a_1^k\rangle+
\langle (a_2^\dagger)^k a_2^k\rangle}{\langle (a_1^\dagger )^k
a_1^k\rangle +
\langle (a_2^\dagger)^k a_2^k\rangle}\right)^2\\
\nonumber &+ 4 \frac{\left|\langle (a_1^\dagger)^k
a_2^k\rangle\right|^2-\langle (a_1^\dagger)^k a_1^k\rangle \langle
(a_2^\dagger)^k a_2^k\rangle}{\left(\langle (a_1^\dagger )^k
a_1^k\rangle + \langle (a_2^\dagger)^k a_2^k\rangle \right)^2}.
\end{align}
The Cauchy-Schwarz inequality predicts
 \begin{align} \label{eq7}
\left|\langle (a_1^\dagger)^k a_2^k\rangle\right|^2 \leq \langle
(a_1^\dagger)^k a_1^k\rangle \langle (a_2^\dagger)^k a_2^k\rangle.
\end{align}
Therefore, the second term in Eq. (\ref{eq6}) is negative or zero.
This result leads to the inequality for higher order duality,
\begin{align} \label{eq8}
D_k^2+V_k^2 \leq 1,
\end{align}
which is the main conclusion in this paper. Although this inequality
has a similar formula to Eq. (\ref{eq1}), it undoubtedly carries
more information about the wave-particle duality and helps deepen
our understanding. In fact, we have generalized the duality relation
from single-photon fields to multi-photon fields. In a typical
duality experiment, if the interferometer (see Fig. 1) is fed with
multi-photons, besides single-photons, according to our conclusion,
the fields in the two paths have to obey not only the first order
duality relation (\ref{eq1}), but also the higher order duality
relation (\ref{eq8}). Since $n$-photon component in a field only
contributes the $k$th-order correlation function and the $k$th-order
coherence with $k\le n$, and takes no effect for the
$k^{\prime}$th-order correlation function or coherence if
$k^{\prime} > n$, all higher than $n$th-order duality information
does not exist (sums up to zero) for the case that at most
$n$-photon component is found in the field. As a consequence, for a
state with $n$-photons there exist $n$ inequalities, $D_k^2+V_k^2
\leq 1$, with $k=1,2, \dots, n$. That is why we only need to
consider the first order distinguishability (\ref{eq2}) and the
first order visibility (\ref{eq3}) in the duality experiments with
single-photons.

\begin{figure}
\includegraphics[width=0.4\textwidth]{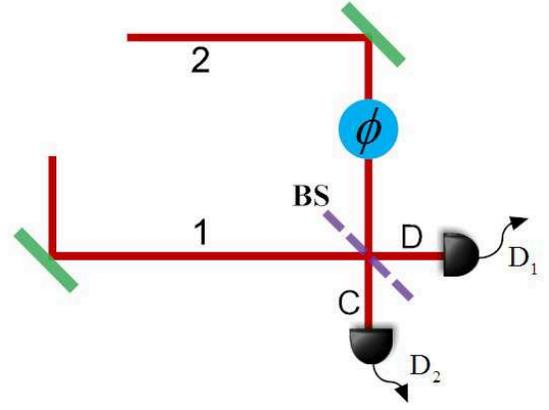}
\caption{The distinguishability and visibility are measured in the
open (removing the beam splitter BS) and closed (employing the beam
splitter BS) interferometer, respectively.}
\end{figure}

From Eq. (\ref{eq6}), it is easy to find that the equality sign in
the higher order duality relation (\ref{eq8}) will be satisfied
under the condition $|\langle (a_1^\dagger)^k
a_2^k\rangle|^2=\langle (a_1^\dagger)^k a_1^k\rangle \langle
(a_2^\dagger)^k a_2^k\rangle$. For example, the $k$th-order duality,
if it exists, is saturated for $k$-photon pure states, which is a
generalization of the first order duality relation $D_1^2+V_1^2 =1$
for the single-photons in a pure state. The condition on which the
$k$th-order duality (\ref{eq8}) achieves its maximum value unity is
usually very complicated if $n$-photon component with $n>k$ is
involved in the field, and $|\langle (a_1^\dagger)^k
a_2^k\rangle|^2=\langle (a_1^\dagger)^k a_1^k\rangle \langle
(a_2^\dagger)^k a_2^k\rangle$ is at present the only test equation
we can obtain.


\section{Physical interpretation}
The $k$th-order auto-correlation $(a_i^\dagger )^k a_i^k$ (i=1,2)
used in the definition (\ref{eq5a}) is equal to,
\begin{align} \label{new1}
(a_i^\dagger )^k a_i^k=\prod_{j=0}^{k-1} (\hat{n}_i^\dagger-j) ,
\end{align}
with the number operator $\hat{n}_i=a_i^\dagger a_i$. Imposing this
operator onto a number state $|n_i>$, we have, $(a_i^\dagger )^k
a_i^k|n_i>=(
\begin{array}{c}  n  \\  k \end{array}  )|n_i>$, with the
binomial coefficient $(
\begin{array}{c}  n  \\  k \end{array} )=\frac{n!}{k!(n-k)!}$. Thus $\langle(a_i^\dagger )^k a_i^k \rangle$
can be regarded as the combination number of picking out $k$
photons, disregarding order, in path $i$, no matter what state is
prepared for the optical field in this path. The $k$th order
distinguishability $D_k$ then has a very clear physical meaning,
i.e., the normalized difference between the $k$-combinations of the
photons in paths $1$ and $2$.

In the basis $|0_10_2>, |0_11_2>, \cdots, |n_1n_2>$, a general
quantum state for the photons in the interferometer can be described
by a density matrix,
\begin{align}
\rho=\left (\begin{array}{ccc}  \rho_{11} ~& \cdots & \rho_{1(n+1)^2}  \\
\vdots ~& \ddots  & \vdots  \\
\rho_{(n+1)^21} ~& \cdots  & \rho_{(n+1)^2(n+1)^2}
\end{array} \right )
\end{align}
Under this quantum state, we can directly write down the expectation
value of the $k$th order distinguishability,
\begin{align}
\langle D_k\rangle= \frac{\sum_{i=k}^n\sum_{j=0}^n\left
(\rho_{p_{i,j}p_{i,j}}-\rho_{q_{i,j}q_{i,j}}\right)\left (
\begin{array}{c}  i  \\  k \end{array}
\right)}{\sum_{i=k}^n\sum_{j=0}^n\left
(\rho_{p_{i,j}p_{i,j}}+\rho_{q_{i,j}q_{i,j}}\right)\left (
\begin{array}{c}  i  \\  k \end{array}
\right)},
\end{align}
with $p_{i,j}=i*n+i+j+1$ and $q_{i,j}=j*n+i+j+1$. Here we see that
only the diagonal elements contribute to the distinguishability.

Just as we already mentioned, the visibility in a duality experiment
is actually related to the coherence between the two paths. Thus the
visibility is determined by the off diagonal elements of the density
matrix for the photons in an interferometer. For example, for the
operator $(a_1^\dagger)^k a_2^k \E^{\I k\phi} + (a_2^\dagger)^k
a_1^k\E^{-\I k\phi} $ used in the definition of the $k$th order
visibility (\ref{eq5b}), the off diagonal elements
$<m^{\prime}_1m^{\prime\prime}_2|\rho|(m^{\prime}+k)_1(m^{\prime\prime}-k)_2>$
and
$<m^{\prime}_1m^{\prime\prime}_2|\rho|(m^{\prime}-k)_1(m^{\prime\prime}+k)_2>$
with $m^{\prime},m^{\prime\prime}\in[0,n]$ have contributions. The
diagonal elements play no role in the evaluation of the visibility.
However, for a density matrix, what we can directly measure in
experiments are just the diagonal elements, usually represented as
the photon counting or higher order coincidence counting. So we have
to turn the information carried by the off diagonal elements to
diagonal elements. That is why a $50:50$ beam splitter is to be
employed at the output of the interferometer for the measurement of
the visibility. For more details on the measurement of higher order
visibility, please see section
\uppercase\expandafter{\romannumeral5}.

Now we can conclude that the distribution of the photons in an
interferometer projected on the Fock states, i.e., the diagonal
elements of the density matrix in the Fock state basis, determines
the photons' particle information, and the wave information relies
on the off diagonal elements, no matter what order
distinguishability and visibility is considered.


\section{Measurement scheme for the visibility}
Compared with the measurement of the distinguishability $D_k$, which
is directly defined by the auto-correlation, the measurement of the
fringe visibility $V_k$, which depends on higher order coherence, is
more complicated. For $k=1$, there exist a straight forward way to
measure $V_1$. Suppose the beam splitter in Fig. 1 is active and the
modes after the beam splitter are denoted with $C$ and $D$, then the
annihilation operators $c$ and $d$ of the modes in these two paths
are connected to the annihilation operators $a_1$ and $a_2$ of the
modes in paths $1$ and $2$ through the relations,
$c=\frac{1}{\sqrt{2}}(a_1+a_2e^{i\phi})$ and
$d=\frac{1}{\sqrt{2}}(a_1-a_2e^{i\phi})$, respectively, where the
phase difference $\phi$ between the two paths can be controlled in
experiments by a phase shifter (see Fig. 1). The counting difference
between the two detectors $D_1$ and $D_2$ (see Fig. 1) is equal to,
\begin{align} \label{eq9}
\langle c^\dagger c - d^\dagger d \rangle_\phi= \langle a_1^\dagger
a_2 \E^{\I\phi} + a_2^\dagger a_1 \E^{-\I\phi}\rangle,
\end{align}
which is just the first order visibility defined in \eq{eq3}.

The second order correlation function, needed for $V_2$, can also be
measured in experiments based on current technology
\cite{hongoumandel}. However, the measurement of the higher order
visibility $V_k$ is complicated.

For a general description, we now suppose the two detectors, $D_1$
and $D_2$, in Fig. 1 are ideal ones, so that all Fock states with
arbitrary photon number can be directly detected and distinguished.
For the case of $k=2$, we take the sum of the expectation value of
the both detectors and obtain, \BE
\begin{split} \label{eq12}
\langle(c^\dagger)^2 c^2 +(d^\dagger )^2 d^2\rangle_{\phi}
&=\frac{1}{2}\langle (a_1^\dagger )^2 a_1^2 +
(a_2^\dagger )^2 a_2^2 + 4a_1^\dagger a_2^\dagger a_1a_2 \\
&+ (a_1^\dagger )^2 a_2^2 \E^{2\I \phi}+(a_2^\dagger )^2 a_1^2
\E^{-2\I \phi} \rangle.
\end{split}
\EE Further retarding the phase shift $\phi$ between the two path
$1$ and $2$ by the value $\pi/2$, we obtain another similar
relation, \BE
\begin{split} \label{eq13}
\langle(c^\dagger)^2 c^2 +(d^\dagger )^2 d^2\rangle_{\phi+\pi/2}
&=\frac{1}{2}\langle (a_1^\dagger )^2 a_1^2 +
(a_2^\dagger )^2 a_2^2 + 4a_1^\dagger a_2^\dagger a_1a_2 \\
&- (a_1^\dagger )^2 a_2^2 \E^{2\I \phi}-(a_2^\dagger )^2 a_1^2
\E^{-2\I \phi} \rangle.
\end{split} \EE
In the following, we use two symbols $R^{\pm}_{k,\phi}$, whose
values are in principle obtainable in experiments, to replace the
expectation values $\langle(c^\dagger)^k c^k \pm(d^\dagger )^k
d^k\rangle_{\phi}$. Thus the equality (\ref{eq9}) can be rewritten
as $\langle a_1^\dagger a_2 \E^{\I\phi} + a_2^\dagger a_1
\E^{-\I\phi}\rangle=R^{-}_{1,\phi}$, and the quantity
$\langle(a_1^\dagger )^2 a_2^2 \E^{2\I \phi}+(a_2^\dagger )^2 a_1^2
\E^{-2\I \phi} \rangle$, involved in both equalities (\ref{eq12})
and (\ref{eq13}), can be described by \BE \label{add1}
\begin{split}
\langle(a_1^\dagger )^2 a_2^2 \E^{2\I \phi}+(a_2^\dagger )^2 a_1^2
\E^{-2\I \phi} \rangle=R^{+}_{2,\phi}- R^{+}_{2,\phi+\pi/2}.
\end{split}
\EE The maximum value of $\langle(a_1^\dagger )^2 a_2^2 \E^{2\I
\phi}+(a_2^\dagger )^2 a_1^2 \E^{-2\I \phi} \rangle$ over the phase
factor $\phi$, required in the quantification of the second order
visibility $V_2$, can then be evaluated by,
 \BE
\begin{split}
|\langle(a_1^\dagger )^2 & a_2^2 \E^{2\I \phi}+(a_2^\dagger )^2
a_1^2
\E^{-2\I \phi} \rangle|_{\text{max by } \phi}^2= \\
&\left( R^{+}_{2,\phi^\prime}- R^{+}_{2,\phi^\prime+\pi/2}\right)^2
+\left( R^{+}_{2,\phi^\prime-\pi/4}-
R^{+}_{2,\phi^\prime+\pi/4}\right)^2,
\end{split}
\EE where $( R^{+}_{2,\phi^\prime}- R^{+}_{2,\phi^\prime+\pi/2})$ is
the real part of the vector $2\langle(a_1^\dagger )^2 a_2^2\E^{2\I
\phi^\prime}\rangle$ (see Eq. (\ref{add1})), and $(
R^{+}_{2,\phi^\prime-\pi/4}- R^{+}_{2,\phi^\prime+\pi/4})$ is the
real part of the vector $2\langle(a_1^\dagger )^2 a_2^2\E^{2\I
(\phi^\prime-\pi/4)}\rangle$, which is equal to the imaginary part
of the vector $2\langle(a_1^\dagger )^2 a_2^2\E^{2\I
\phi^\prime}\rangle$. The absolute value of this quantity is
mathematically equivalent to the unnormalized visibility $V_2$, due
to the relation $|\langle(a_1^\dagger )^k a_2^k \E^{\I
k\phi}+(a_2^\dagger )^k a_1^k \E^{-\I k\phi} \rangle|_{\text{max by
} \phi}=2|\langle(a_1^\dagger )^k a_2^k\rangle|$. The phase
$\phi^\prime$ can be arbitrarily chosen, because the modulus of a
vector should remain invariant under the rotation of the coordinate
system.

In general, the term $\langle(a_1^\dagger )^k  a_2^k \E^{\I
k\phi}+(a_2^\dagger )^k a_1^k \E^{-\I k\phi} \rangle$ can be
determined by adding and subtracting $\langle (c^\dagger)^k
c^k\pm(d^\dagger)^k d^k\rangle_\phi$ for $k$ different values of
$\phi$. For example, for odd number of $k$, we have \BE
\begin{split}
\langle(a_1^\dagger )^k & a_2^k \E^{\I k\phi}+(a_2^\dagger )^k a_1^k
\E^{-\I k\phi} \rangle=
\frac{2^{k-1}}{k}\sum_{m=0}^{k-1}R^-_{k,\phi+2m\pi/k}.
\end{split}
\EE Accordingly, the maximum value of $\langle(a_1^\dagger )^k a_2^k
\E^{\I k\phi}+(a_2^\dagger )^k a_1^k \E^{-\I k\phi} \rangle$ over
the phase factor $\phi$, used for the $k$th order visibility $V_k$,
can be evaluated by,
 \BE
\begin{split}
|\langle(&a_1^\dagger )^k  a_2^k \E^{\I k\phi}+(a_2^\dagger )^k
a_1^k
\E^{-\I k\phi} \rangle|_{\text{max by }\phi}^2= \left(\frac{2^{k-1}}{k}\right)^2\times\\
&\left[\left(\sum_{m=0}^{k-1}R^-_{k,\phi^\prime+2m\pi/k}\right)^2+\left(\sum_{m=0}^{k-1}R^-_{k,\phi^\prime-\pi/(2k)+2m\pi/k}\right)^2\right],
\end{split}
\EE where the phase $\phi^\prime$ can be arbitrarily chosen.

For even number of $k$, we have \BE
\begin{split}
\langle(a_1^\dagger )^k & a_2^k \E^{\I k\phi}+(a_2^\dagger )^k a_1^k
\E^{-\I k\phi} \rangle=
\frac{2^{k-1}}{k}\sum_{m=0}^{k-1}(-1)^mR^+_{k,\phi+m\pi/k}.
\end{split}
\EE The maximum value of $\langle(a_1^\dagger )^k a_2^k \E^{\I
k\phi}+(a_2^\dagger )^k a_1^k \E^{-\I k\phi} \rangle$ over the phase
factor $\phi$, used for the $k$th order visibility $V_k$, can be
evaluated by,
 \BE
\begin{split}
&|\langle(a_1^\dagger )^k  a_2^k \E^{\I k\phi}+(a_2^\dagger )^k
a_1^k
\E^{-\I k\phi} \rangle|_{\text{max by } \phi}^2= \left(\frac{2^{k-1}}{k}\right)^2\times\\
&\left[\left(\sum_{m=0}^{k-1}(-1)^mR^+_{k,\phi^\prime+m\pi/k}\right)^2+\left(\sum_{m=0}^{k-1}(-1)^mR^+_{k,\phi^\prime-\pi/(2k)+m\pi/k}\right)^2\right],
\end{split}
\EE with an arbitrary phase factor $\phi^\prime$.

\section{Conclusions}
The distinguishability of photons' passage in the interferometer and
the visibility of the interference pattern after the interferometer
can be regarded as the first order particle-like information and the
first order wave-like information. By introducing the concepts of
higher order distinguishability and visibility for multi-photons,
which are related to higher order auto-correlation and coherence
between the fields in the two paths of the interferometer, we
generalize the wave-particle duality relation from the first order
case to higher order case. We believe it to be a useful tool for
analyzing the duality experiments with the input of multi-photons,
or even a classical light. If we do the duality experiment by using
different light sources, the same results may be obtained if only
the first order duality is considered. However, we believe it will
exhibit different results for higher order duality information. The
concept of higher order duality may provide us more information
about the duality experiments, especially with the input of
multi-photons, and accordingly helps us deepen the understanding on
the wave-particle duality.

\section*{Acknowledgements}
This research was supported by the National Basic Research Program
of China (Grant No. 2011CB922203 and 2012CB921603), the national
Natural Science Foundation of China (Grant No. 11174118 and
11174026), and the Natural Science Foundation of Jiangxi Province
under Grant No. 20114BAB212003. The research of MSZ is supported by
NPRP grant (No. 4-346-1-061) from Qatar National Research Fund.


\end{document}